\documentstyle[12pt]{article}

\def\ni{\noindent}

\def\del{\partial}
\def\dis{\displaystyle}

\begin{document}

\begin{center}
{\large\bf On the Canonical Formalism for a Higher-Curvature Gravity}\\[8mm]
Yasuo Ezawa${}^1$, Masahiro Kajihara${}^1$, Masahiko Kiminami${}^1$, 
Jiro Soda${}^2$ and Tadashi Yano${}^3$\\[3mm]
${}^1$ Department of Physics, Ehime University\\
 Mtsuyama 790 Japan\\
${}^2$ Department of Fundamental Sciences, FIHS
Kyoto University, Kyoto 606 Japan\\
${}^3$ Department of Electrical and Electronic Engineering, Ehime University\\
Matsuyama, 790 Japan\\[1cm]
\end{center}

\ni
{\bf Abstract}

Following to the method of Buchbinder and Lyahovich, we carry out a canonical 
formalism for a higher-curvature gravity theory in which the Lagrangian density
${\cal L}$ is given in terms of a function of the scalar curvature $R$ as 
${\cal L}=\sqrt{-\det g_{\mu\nu}}f(R)$. 
The local Hamiltonian is obtained by a canonical transformation which 
interchanges a pair of the generalized coordinate and their canonical momentum
coming from the time derivative of the metric.
\\[1.5cm]

\ni
{\bf \S 1. Introduction}\\

The higher curvature gravity (HCG) theory is derived naturally from 
the superstring theory as its low energy effective theory.
HCG is one of the modifications of the Einstein gravity.
For macroscopic phenomena, the Einstein gravity works sufficiently well
so that the modification is negligible.
The modification is, however, expected to play an important role in 
microscopic, i.e. Planck scale phenomena.
In most of the unified theories of fundamental physics, such as superstring 
, supergravity theory and Kaluza-Klein theories, etc. extra-dimensional 
spaces are necessary.
The extra-dimensional spaces are expected to have contracted to Planck scale
during inflation.
Then there arises the problem of the stabilization of the extra-dimensional 
spaces which has been investigated by many authors 
\cite{CHPS,Maeda,AGHK,B-Z,I-M}.
In the Einstein gravity, it seems to be difficult to stabilize 
the extra-dimensional spaces after inflation \cite{ESWY,EKSY}.
So it is natural to ask whether these spaces are stabilized in HCG.

If HCG plays a crucial role in the stabilization, the mechanism would be 
quantum mechanical.
Then we are lead to the canonical formalism of HCG which has been investigated 
by Boulware\cite{Boul} and by Buchbinder \& Lyahovich (BL)\cite{BL}.
The latter authors generalize the formalism of the higher-derivative theory by 
Ostrogradski to an appropriate form for HCG.
They modify the Lagrangian density, (2.1) given below, by using the Lagrange 
multiplier method which takes into account the definitions of new coordinates 
and their relations to the time derivatives of the original coordinates.
The advantages of the method are that, once the modified Lagrangian density 
is constructed, all the equations are derived by the variational principle and 
the consistency conditions, and that it is not necessary that the orders of 
the highest time derivatives of the coordinates are the same.
Thus the procedure is straightforward although tedious.
They found that, if the additional term in the modified Lagrangian contains 
only the square of the scalar curvature, there exist second class constraints.
However, they have not obtained the Lagrange multipliers and the final form of 
the Hamiltonian constraint.

In the formalism of Boulware, the Lagrangian density is restricted to 
the case in which it contains terms constructed from the Ricci tensor. 
The generalization to the Lagrangian density given by (2.1) is not 
straightforward\cite{BKL,Kasp}.

In this work we will complete the formalism of BL.
We reduce the phase space by using the second class constraints.
We have only one remaining coordinate arising from the time derivative of 
the metric\cite{BL}.
The last step is the canonical transformation which interchanges the new 
generalized coordinate and its conjugate mementum.
This transformation avoids the non-locality when we proceed to 
the quantum theory.
The  Hamiltonian thus obtained has more natural dependence on the coordinate 
and momentum than before the transformation.
We will also comment on the relation to the so called equivalence theorem 
\cite{Whitt,MaedaII} well known in classical HCG .

In section 2, we develope the canonical formalism for the HCG.
In section 3, the formalism is applied to the Robertson-Walker spacetime.
Section 4 is devoted to the summary and discussion.
The problem of stabilization of the extra-dimensional spaces will be 
investigated in a separate paper.
\\

\ni
{\bf \S2. Canonical Formalism}\\

In this section we present a canonical formalism of a HCG whose Lagrangian 
density is given by
$$
     {\cal L}=\sqrt{-\det g_{\mu\nu}}f(R)                          \eqno(2.1)
$$
where $f(R)$ is a function of the scalar curvature $R$.
In terms of the ADM variables\cite{ADM}, the three dimensional metric 
$\;h_{ij}\equiv g_{ij}\;$, the lapse function $\,N\,$ and the shift vector 
$\,N^i\,$, the scalar curvature $\,R\,$ is expressed in the following $d+1$ 
form: 
$$
     R=2g^{mn}{\cal L}_{n}K_{mn}+{\bf K}^2-3K_{mn}K^{mn}+{\bf R}   
      -2N^{-1}\Delta N.                                            \eqno(2.2)
$$
Here $K_{kl}$ is the extrinsic curvature
$$
     K_{kl}\equiv{1\over2}{\cal L}_{n}h_{kl}
                  ={1\over 2N}(\del_{0}h_{kl}-N_{k;l}-N_{l;k}),    \eqno(2.3)
$$
$n^{\mu}$ is the normal to the surface $\,t={\rm constant}\,$
$$
     n^{\mu}=N^{-1}(1,-N^m).                                       \eqno(2.4)
$$
and ${\cal L}_{n}$ is the Lie derivative with respect to the normal $n^{\mu}$
$$
     {\cal L}_{n}K_{kl}=N^{-1}(\del_{0}K_{kl}-N^jK_{kl;j}
                       -N^j_{\ ;k}K_{jl}-N^j_{\ ;l}K_{kj}).        \eqno(2.5)
$$
In the above (and hereafter), a bold face letter denotes the trace of a 2nd 
rank tensor, e.g. $h_{kl}R^{kl}\equiv{\bf R}\;$.

Now following BL, we introduce new generalized coordinates
$$
     Q_{ij}\equiv K_{ij}                                           \eqno(2.6)
$$
and $``$velocity" variables
$$
     v_{ij}\equiv\del_{0}Q_{ij},\ \ \ v^i\equiv\del_{0}N^i,
             \ \ \ {\rm and} \ \ \ v\equiv\del_{0}N.               \eqno(2.7)
$$
Then there are relations between the new coordinates and the time derivatives 
of the original coordinates
$$
     \del_{0}h_{ij}=2NQ_{ij}+N_{i;j}+N_{j;i}.                      \eqno(2.8)
$$
In terms of the new variables, Lagrangian density (denoted as ${\cal L}_{Q}$)
is expressed as
$$
     {\cal L}_{Q}=\sqrt{\det h_{ij}}Nf(R_{Q}).                     \eqno(2.9)
$$
where
$$
     R_{Q}=2N^{-1}h^{kl}v_{kl}-{U}_{Q}-2N^{-1}\Delta N
          -2N^{-1}(N^k{\bf Q}_{;k}+2N^{k;l}Q_{kl})                 \eqno(2.10)
$$
and 
$$
     {U}_{Q}=3Q_{mn}Q^{mn}-{\bf Q}^2-{\bf R}.      \eqno(2.11)
$$
Using  the method of Lagrange multiplier, modified Lagrangian density 
${\cal L}^{*}$, which takes into account the relations (2.8) and 
the definitions (2.6) and (2.7), is given as follows:
$$
\begin{array}{ll}
     {\cal L}^{*}&={\cal L}_{Q}+p^{kl}\{\del_{0}h_{kl}-(2NQ_{kl}
                                                      +N_{k;l}+N_{l;k})\}
\\[3mm]
                 &\hspace{5mm}+P^{kl}(\del_{0}Q_{kl}-v_{kl})
                              +P_{k}(\del_{0}N^k-v^k)+P(\del_{0}N-v).
                  \end{array}                                      \eqno(2.12)
$$
The Hamiltonian density ${\cal H}^{*}$ corresponding to ${\cal L}^{*}$ is 
given as usual
$$
\begin{array}{ll}
     {\cal H}^*&\equiv P^{kl}\del_{0}{Q}_{kl}+P_{k}\del_{0}N^k+P\del_{0}N
                 +p^{kl}\del_{0}h_{kl}-{\cal L}^*
\\[3mm]
               &=p^{kl}(2NQ_{kl}+N_{k;l}+N_{l;k})+P^{kl}v_{kl}+P_{k}v^k+Pv
                 -{\cal L}_{Q}.
\end{array}                                                        \eqno(2.13)
$$
The action $S$ to be used in the variational principle is given by
$$
     S=\int d^{d+1}x\,{\cal L}^{*}                                 \eqno(2.14)
$$
The variations of the velocitry variables give the $``$momenta" 
$P^{kl},\;P_{k}\ {\rm and}\ P$ which are expressed as\footnote{
In the Ostrogradski's formalism, these equations define the momenta canonically
 conjugate to the highest time derivatives of the generalized coordinates.}
$$
\left\{\begin{array}{l}
     \dis P^{kl}={\del{\cal L}_{Q}\over \del v_{kl}}
                =2\sqrt{h}f'(R_{Q})h^{kl}
\\[5mm]
     \dis P_{k}={\del{\cal L}_{Q}\over\del v^k}=0
\\[5mm]
     \dis P={\del{\cal L}_{Q}\over \del v}=0
\end{array}\right.                                                 \eqno(2.15)
$$
where $f'=df/dR\,$ and $\,h=\det h_{ij}\,$.
The velocity variables in (2.13) should be replaced by these momenta.
Equation (2.15), however, cannot be solved for all velocity variables, i.e. 
there are primary constraints.
Decomposing the second rank tensor (density) into a traceless part (denoted by 
a tilde) and a trace part
$$
     T_{kl}=\tilde{T}_{kl}+{1\over d}h_{kl}{\bf T},                \eqno(2.16)
$$ 
we obtain the following expressions for these constraints
$$
\left\{\begin{array}{l}
     \dis C^{kl}\equiv \tilde{P}^{kl} \approx 0
\\[5mm]
     C_{k}\equiv P_{k}\approx 0
\\[5mm]
     C\equiv P\approx 0
\end{array}\right.                                                 \eqno(2.17)
$$
where $k,l$ run from 1 to $d$  and $\approx 0$ means that the quantity vanishes
due to the constraints.
The only velocity variable that can be solved from (2.15) is the trace of 
$v_{kl}$
$$
     {\bf v}={N\over2}[\psi({\bf P}/2d\sqrt{h})+U_{Q}]+\Delta N
             +N^k\del_{k}{\bf Q}+2N^{k;l}Q_{kl}.                   \eqno(2.18)
$$
where $\,{\bf P}\,$, the trace part of $P^{kl}$, is given as
$$
     {\bf P}=2d\sqrt{h}f'(R_{Q})\ \ \ {\rm or}\ \ \ 
     R_{Q}=f'^{-1}({\bf P}/2d\sqrt{h})\equiv\psi({\bf P}/2d\sqrt{h})                                                                               \eqno(2.19)
$$
Using  (2.19), we can express ${\cal H}^*\;$  as
$$
\begin{array}{ll}
     {\cal H}^*&=\dis C^{kl}\tilde{v}_{kl}+C_{k}v^k+Cv
\\[3mm]
               &\dis +N\Bigl[\;2p^{kl}Q_{kl}
                     +{1\over2d}{\bf P}\psi({\bf P}/2d\sqrt{h})
                     +{1\over 2d}{\bf P}U_{Q}
                     -\sqrt{h}f(\psi({\bf P}/2d\sqrt{h}))\;\Bigr]
                     +{1\over d}{\bf P}\Delta N
\\[3mm]
                &\dis +2p^{kl}N_{k;l}+{1\over d}{\bf P}(N^k\del_{k}{\bf Q}
                                                       +2N^{k;l}Q_{kl}).
\end{array}                                                        \eqno(2.20)
$$
The canonical Hamiltonian density ${\cal H}_{0}$ is obtained by putting 
$$
    {\cal H}^*={\cal H}_{0}+C^{kl}\tilde{v}_{kl}+C_{k}v^k+Cv.      \eqno(2.21)
$$
The velocity variables on the right hand side  cannot be expressed in terms of 
the momentum variables and play the role of the Lagrange multipliers.
${\cal H}_{0}$ is written in the form
$$
     {\cal H}_{0}=N{\cal H}+N^k{\cal H}_{k}+{\rm divergent\ terms}.\eqno(2.22)
$$
Here
$$
\left\{\begin{array}{ll}
     {\cal H}&\dis=2Q_{kl}p^{kl}+{1\over2d}{\bf P}[\psi({\bf P}/2d\sqrt{h})
                                  +U_{Q}]+{1\over d}\Delta {\bf P}
                                  -\sqrt{h}f(\psi({\bf P}/2d\sqrt{h})
\\[5mm]
     {\cal H}_{k}&\dis=-2h_{km}p^{mn}_{\ \ ;n}
                       +{1\over d}\Bigl[{\bf P}\del_{k}{\bf Q}
                                        -2({\bf P}Q_{kl})^{;l}\Bigr].
\end{array}\right.                                                 \eqno(2.23)
$$
The Poisson bracket(PB) form of the canonical equations of motion are derived 
from  $\;{\cal H}^{*}\;$ if we treat the pairs of the generalized 
coordinates and the Lagrange multipliers, $\;(P^{kl},Q_{kl})$,
$\;(p^{kl},h_{kl}), \;(P,N), \;(P_{k},N^k)\;$, as canonically conjugate pairs.
Then it is easily seen that the Poisson brackets(PB's) among the primary 
constraints vanish.
Furthermore PB's of $N^k{\cal H}_{k}$ and the primary constraints also vanish.
This means that the time variation of the primary constraints are solely given 
by $N{\cal H}$.
Thus the conditions that the primary constraints should be preserved are that 
the PB's of $N{\cal H}$ and the primary constraints should  vanish and are 
expressed by the following set of secondary constraints as
$$
\left\{\begin{array}{l}
     {\cal C}\equiv{\cal H} \approx 0 
\\[3mm]
     {\cal C}_{k}\equiv{\cal H}_{k} \approx 0 
\\[5mm]
     \dis{\cal C}^{kl}\equiv  p^{kl}-{1\over d}{\bf p}h^{kl}
                      +{1\over 2d}{\bf P}\Bigl(Q^{kl}-{1\over d}h^{kl}{\bf Q}
                        \Bigr)\approx 0.
\end{array}\right.                                                 \eqno(2.24)
$$
The first and the second constraints are interpreted as the $``$Hamiltonian" 
constraint and the $``$momentum" constraint, respectively.

The PB's of the Hamiltonian constraint and the primary constraints also 
vanishes owing to the secondary constraints.
The PB's of the momentum constraint and the primary constraints vanish,
whch is desirable if the former one is the generator of the spacial coordinate
transformations, which is to be seen.
The third  one gives the momenta canonically conjugate to $h_{kl}$ which is 
not given by the equation like (2.13) but was introduced as a Lagrange 
multiplier.
Only non-vanishing PB's among the primary  constraints and $\;{\cal C}^{kl}\;$
is the following:
$$
     \{C^{kl}({\bf x}),{\cal C}^{ij}({\bf x}')\}_{PB}
      ={1\over2d}{\bf P}(h^{ki}h^{lj}-{1\over d}h^{kl}h^{ij})
        \delta({\bf x}-{\bf x}').                                  \eqno(2.25)
$$
The constraints $C,\ C_{k}$ are first class and $\;C^{kl},\ {\cal C}^{kl}
\;$ are second class constraints\cite{BL}.
Thus the velocity variables $\tilde{v}_{kl}$ are determined from the condition 
that the secondary constraints ${\cal C}^{kl}\approx 0$ should be preserved in 
time:

$$
\begin{array}{ll}
     \tilde v_{kl}&\dis=2N\Bigl[\,({\bf P}^{-1}{\bf p}
                       -{d-1\over 2d}{\bf Q})\tilde{Q}_{kl}
                       +Q_{km}Q_{l}^{\ m}-{1\over d}h_{kl}Q_{mn}Q^{mn}
                      -{1\over2}\tilde{R}_{kl}\,\Bigr]
\\[5mm]
                   &\dis\hspace{5mm}-{\bf P}^{-1}(\nabla_{k}\nabla_{l}
                       -{1\over d}h_{kl}\Delta)(N{\bf P})
\\[5mm]
                   &\dis+\tilde{Q}_{km}N^m_{\ ;l}+\tilde{Q}_{lm}N^m_{\ ;k}
                        -(\tilde{Q}_{k}^{\ m}N_{l;m}+\tilde{Q}_{l}^mN_{k_;m})
                        +{\bf P}^{-1}({\bf P}N^j)_{;j}\tilde{Q}_{kl}
\\[5mm]
                   &\dis+2{\bf P}^{-1}({\bf p}+{1\over 2d}{\bf PQ})
                             (N_{k;l}+N_{l;k}-{2\over d}h_{kl}N^j_{\ ;j}).
\end{array}                                                        \eqno(2.26)
$$
Apart from divergent terms, we have
$$
\begin{array}{ll}
     C^{kl}\tilde{v}_{kl}&\dis=2N\Bigl[\,({\bf P}^{-1}{\bf p}
                             -{d-1\over 2d}{\bf Q})\tilde{P}^{kl}Q_{kl}
                             +{1\over2}{\bf P}({\bf P}^{-1}P^{kl})_{;kl}
                             +\tilde{P}^{kl}Q_{km}Q_{l}^{\ m}
                             -{1\over2}\tilde{P}^{kl}R_{kl}
\\[5mm]
                        &\dis+N^k\Bigl[\,2\{(\tilde{P}^m_{\ k}Q_{m}^{\ n})_{;n}
                                             -(\tilde{P}^{mn}Q_{mk})_{;n}\}
                            +{\bf P}({\bf P}^{-1}\tilde{P}^{mn}Q_{mn})_{;k}
\\[5mm]
                        &\dis\hspace{1cm}+4\{{\bf P}^{-1}({\bf p}
                            +{1\over2d}{\bf PQ})\tilde{P}^l_{\ k}\}_{;l}
                              \,\Bigr]
\end{array}                                                        \eqno(2.27)
$$
Then we obtain the following form for ${\cal H}^{*}$
$$
     {\cal H}^{*}=C_{k}v^k+Cv+N\bar{{\cal H}}+N^k\bar{{\cal H}_{k}}
                   +{\rm divergent\ terms}                         \eqno(2.28)
$$
where
$$
\left\{\begin{array}{ll}
    \bar{{\cal H}}&=\dis 2p^{kl}Q_{kl}+{1\over 2d}{\bf P}\psi-\sqrt{h}f(\psi)
                         
                    +{\bf P}^{-1}({\bf p}+{5-d\over 2d}{\bf PQ})
                                          \tilde{P}^{kl}\tilde{Q}_{kl}
                    +2\tilde{P}^{kl}\tilde{Q}_{k}^{\ m}\tilde{Q}_{lm}
\\[5mm]
                 &\dis+{1\over 2d}{\bf P}(3\tilde{Q}_{mn}\tilde{Q}^{mn}
                            +{3-d\over d}{\bf Q}^2)
                      +{\bf P}({\bf P}^{-1}\tilde{P}^{kl})_{;kl}
                      +{1\over d}\Delta {\bf P}
                      -(\tilde{P}^{kl}\tilde{R}_{kl}+{1\over2d}{\bf PR})
\\[5mm]
     \bar{{\cal H}}_{k}&\dis=-2h_{km}p^{mn}_{\ \ ;n}
                       +{1\over d}\Bigl[{\bf P}\del_{k}{\bf Q}
                                        -2({\bf P}Q_{kl})^{;l}\Bigr]
                       +{\bf P}({\bf P}^{-1}\tilde{P}^{mn}\tilde{Q}_{mn})_{;k}
\\[5mm]
                 &\dis+2[(\tilde{P}^l_{\ k}Q_{l}^{\ m})_{;m}
                  -(\tilde{P}^{ml}\tilde{Q}_{mk})_{;l}]
                  +4[{\bf P}^{-1}({\bf p}+{1\over2d}{\bf PQ})
                                         \tilde{P}_{k}^{\ l}]_{;l}
\end{array}\right.                                                 \eqno(2.29)
$$
It appears from eq.(2.29) that it would be more convenient to choose 
$\tilde{Q}_{ij}$ and ${\bf Q}$ as generalized coordinates than $Q_{ij}$.
Since we have
$$
     P^{ij}\dot{Q}_{ij}=\tilde{P}^{ij}\dot{\tilde{Q}}_{ij}
                       +{1\over d}{\bf P}\dot{\bf Q}
                       +{1\over d}({\bf P}{Q}^{ij}
                                          +\tilde{P}^{ij}{\bf Q})\dot{h}_{ij},
$$
corresponding canonical momenta are $\;\tilde{P}^{ij},\ {\bf P}/d\equiv 
\hat{{\bf P}}\;$ and $p'^{ij}\;$ which is given by
$$
      p'^{ij}=p^{ij}+\hat{{\bf P}}Q^{ij}+{1\over d}\tilde{P}^{ij}{\bf Q}
             =p^{ij}+\hat{{\bf P}}(\tilde{Q}^{ij}+{1\over d}h^{ij}{\bf Q})
                    +{1\over d}\tilde{P}^{ij}{\bf Q}.
                                                                   \eqno(2.30)
$$
It is easily seen that these variables obey the usual Poisson brackets among 
the canonical pairs of variables.
Then $\,\bar{\cal H}\;$ and $\,\bar{\cal H}_{k}\,$ in (2.29) are expressed as
$$
\left\{\begin{array}{ll}
    \bar{{\cal H}}&=\dis 2p'^{ij}(\tilde{Q}_{ij}+{1\over d}h_{ij}{\bf Q})
                   +{1\over 2}\hat{\bf P}\psi-\sqrt{h}f(\psi)
\\[5mm]
                   &\dis -{1\over 2}\hat{\bf P}\tilde{Q}^{ij}\tilde{Q}_{ij}
                   -{d+1\over 2d}\hat{\bf P}{\bf Q}^2+\Delta \hat{\bf P}
                   -{1\over 2}\hat{\bf P}{\bf R}
\\[5mm]
                   &\dis +{1\over d}\hat{\bf P}^{-1}{\bf p}'\tilde{P}^{ij}
                          \tilde{Q}_{ij}
                   -{d+1\over 2d}{\bf Q}\tilde{P}^{ij}\tilde{Q}_{ij}
                   +2\tilde{P}^{ij}\tilde{Q}_{i}^{\ l}\tilde{Q}_{jl}
                   +\hat{\bf P}(\hat{\bf P}^{-1}\tilde{P}^{ij})_{;ij}
                   -\tilde{P}^{kl}\tilde{R}_{kl}
\\[5mm]
     \bar{{\cal H}}_{k}&=\dis-2h_{km}p'^{mn}_{\ \ ;n}
                             +\hat{\bf P}{\bf Q}_{;k}
\\[5mm]
                  &\dis +\hat{\bf P}(\hat{\bf P}^{-1}\tilde{P}^{mn}
                         \tilde{Q}_{mn})_{;k}
                        +2(\tilde{P}_{k}^{\ l}\tilde{Q}_{l}^{\ m}
                        -\tilde{P}^{lm}\tilde{Q}_{kl})_{;m}
                        +{4\over d}(\hat{\bf P}^{-1}{\bf p}'
                           \tilde{P}_{k}^{\ l})_{;l}
\end{array}\right.                                                 \eqno(2.31)
$$

To proceed further, one of the two methods is usually adopted.
One is to use only independent variables by solving the constraint equations .
The other is to use the Dirac brackets without solving the constraints, which 
is sometomes more convenient when we cosider symmetry properties.
Here we adopt the former method since it is simpler.
As to the latter one, we only present the list of the Dirac brackets in 
the appendix.

Now we reduce the phase space by using the second class constraints, 
$\tilde{P}^{ij}\approx 0\;$ and $\;\tilde{Q}^{ij}\approx 
2\hat{\bf P}^{-1}\tilde{p}'^{ij}\;$.
Remaining independent canonical pairs of variables are $(h_{ij},p'^{ij})$ and 
$({\bf Q},\hat{\bf P})$, in terms of which $\,\bar{\cal H}\;$ and 
$\,\bar{\cal H}_{k}\,$ in (2.31) are simplified to
$$
\left\{\begin{array}{ll}
    \bar{{\cal H}}&=\dis 2\hat{\bf P}^{-1}(p'^{ij}p'_{ij}-{1\over d}{\bf p}'^2)
                   +{2\over d}{\bf p'Q}+{1\over 2}\hat{\bf P}\psi
                   -\sqrt{h}f(\psi)
\\[5mm]
                   &\dis -{d+1\over 2d}\hat{\bf P}{\bf Q}^2+\Delta \hat{\bf P}
                   -{1\over 2}\hat{\bf P}{\bf R}
\\[5mm]
     \bar{{\cal H}}_{k}&=\dis-2h_{km}p'^{mn}_{\ \ ;n}+\hat{\bf P}{\bf Q}_{;k}
\end{array}\right.                                                 \eqno(2.32)
$$

In classical theory, (2.32) gives the final form for the Hamiltonian constraint
and momentum constraints.
However, when passing to the quantum theory, we meet a difficulty with (2.32)
in that it contains terms with negative power of $\hat{\bf P}$.
Such terms lead to non-locality in the canonical quantum field theory.
Thus we  make a canonical transformation which interchanges the momenta with 
the coordinates
$$
     ({\bf Q},\;\hat{\bf P})\longrightarrow ({\bf Q}',\;\hat{\bf P}')
                                           =(\hat{\bf P},\,-{\bf Q})                                                                                \eqno(2.33)
$$
In terms of the new coordinates and momenta, $\psi$ is a function of 
${\bf Q}'/2\sqrt{h}$ and $\,\bar{{\cal H}},\;\bar{\cal H}_{k}\,$ are 
expressed as
$$
\left\{\begin{array}{ll}
    \bar{{\cal H}}&=\dis 2{\bf Q}^{-1}(p'^{ij}p'_{ij}-{1\over d}{\bf p}'^2)
                   -{2\over d}{\bf p'\hat{P}}+{1\over 2}{\bf Q}\psi
                   -\sqrt{h}f(\psi)
\\[5mm]
                   &\dis -{d+1\over 2d}{\bf Q}\hat{\bf P}^2+\Delta {\bf Q}
                         -{1\over 2}{\bf Q}{\bf R}
\\[5mm]
     \bar{{\cal H}}_{k}&=\dis-2h_{km}p'^{mn}_{\ \ ;n}-{\bf Q\hat{P}}_{;k}
\end{array}\right.                                                 \eqno(2.34)
$$
where we have dropped the primes denoting the new variables for simplicity.
\vspace{8mm}

\noindent
{\bf \S3. Robertson-Walker spacetime}\\

In this section we apply the results of the previous section to 
the Robertson-Walker spacetime whose metric is written as
$$
     ds^2=-N^2(t)dt^2+a^2(t)\Bigl[\,{dr^2\over 1-kr^2}
                                 +r^2(d\theta^2+\sin^2\theta d\phi^2)\,\Bigr].
                                                                   \eqno(3.1)
$$
The scalar curvature $R$ is given by
$$
     R=6N^{-2}\Bigl[\,{\ddot{a}\over a}+\Bigl({\dot{a}\over a}\Bigr)^2
      +{kN^2\over a^2}-{\dot{N}\over N}{\dot{a}\over a}\Bigr]      \eqno(3.2)
$$
The 3-dimensional scalar curvature is $\,{\bf R}=6k/a^2\,$.
The extrinsic curvature $\,K_{ij}\,$ is given as
$$
     K_{ij}={\dot{a}\over Na}\,h_{ij}                              \eqno(3.3)
$$
where $\,h_{ij}\,$ is the 3-dimensional metric tensor and has a form $\ a^2(t)
\times\,$(fixed functions of $\,r,\theta, \phi\,$) as can be seen from (3.1).

The new generalized coordinates $\,Q_{ij}\,$ can be taken to be $\,K_{ij}\,$
as in the previous section:
$$
     Q_{ij}\equiv K_{ij}={\dot{a}\over Na}\,h_{ij}.
                                                                    \eqno(3.4)
$$
Obviously there is only one independent component which we take to be 
the trace part $\,{\bf Q}\,$:
$$
     {\bf Q}=3(\dot{a}/Na)\ \ \ {\rm or}\ \ \ Q_{ij}={1\over3}{\bf Q}h_{ij}.
                                                                   \eqno(3.5)
$$
Thus we take $\,{\bf Q}\,$ to be the new generalized coordinate and 
the corresponding velocity variables as
$$
     {\bf v}\equiv \dot{\bf Q}.                                    \eqno(3.6)
$$
Then the scalar curvature is expressed as
$$
     R=2N^{-1}{\bf v}+{4\over3}{\bf Q}^2+{6k\over a^2}\equiv R_{Q}.                                                                               \eqno(3.7)
$$
Using this $\,R_{Q}\,$, the Lagrangian density is written as
$$
     {\cal L}_{Q}=\sqrt{-g}\,f(R_{Q})                              \eqno(3.8)
$$
The modified Lagrangian density, which takes into account the relation (3.5) 
and the definitions such as (3.6), is given by
$$
     {\cal L}^{*}={\cal L}_{Q}+{\bf P}(\dot{\bf Q}-{\bf v})+P(\dot{N}-v)
                 +p(\dot{a}-{1\over3}Na{\bf Q}).                   \eqno(3.9)
$$
Here ${\bf P}$ was denoted as $\hat{\bf P}$ in the previous section and $p$ is 
related to $p'^{ij}$ as $p=2h_{ij}p'^{ij}/a$.
It is noted that all the fields are homogeneous.
The Hamiltonian density corresponding to this $\,{\cal L}^{*}$ is taken to be
$$
\begin{array}{cl}
     {\cal H}^{*}&\equiv {\bf P}\dot{\bf Q}+P\dot{N}+p\dot{a}-{\cal L}^{*}
\\[3mm]
                 &\dis={\bf P}{\bf v}+Pv+{1\over3}\,Nap{\bf Q}-{\cal L}_{Q}
\end{array}                                                        \eqno(3.10)
$$
Variations with respect to the velocity variables lead to the following 
equations
$$
\begin{array}{l}
     \dis {\bf P}={\delta{\cal L}_{Q}\over \delta {\bf v}}
                 =2\sqrt{h}f'(R_{Q})
\\[5mm]
     \dis P={\delta{\cal L}_{Q}\over \delta v}=0.
\end{array}                                                        \eqno(3.11)
$$
From the first equation of (3.11), we have
$$
     {\bf v}={1\over2}N[\psi({\bf P}/2\sqrt{h})-{4\over3}{\bf Q}^2-6k/a^2]
                                                                   \eqno(3.12)
$$
where
$$
     \psi({\bf P}/2\sqrt{h})\equiv f^{'-1}({\bf P}/2\sqrt{h})=R_{Q}.\eqno(3.13)
$$
The second one of (3.11) gives the primary constraint
$$
     C\equiv P\approx 0                                           \eqno(3.14)
$$
Then $\,{\cal H}^{*}\,$ is written as
$$
     {\cal H}^{*}=Cv+{\bf P}{\bf v}+{1\over3}Na{\bf Q}p-{\cal L}_{Q}                                                                               \eqno(3.15)
$$
where
$$
     {\cal L}_{Q}=N\sqrt{h}f(\psi({\bf P}/6\sqrt{h}).              \eqno(3.16)
$$
Using (3.12) we obtain the following expression for $\,{\cal H}^{*}\,$ 
$$
     {\cal H}^{*}=Cv+N{\cal H}                                     \eqno(3.17)
$$
where
$$
     {\cal H}={1\over 2}{\bf P}(\psi-{4\over3}{\bf Q}^2-6k/a^2)
             +{1\over3}a{\bf Q}p-\sqrt{h}f(\psi)                   \eqno(3.18)
$$
We can treat each pair of variables
$$
     \; (a(t),p(t)),\ \ \ (N(t),P(t)),\ \ \ ({\bf Q}(t),{\bf P}(t))\;,
$$
to be  canonically conjugate as in the previous section and calculate 
the Poisson brackets.

Preservation of the primary constraints yields a secondary constraint
$$
     0\approx \{C,{\cal H}^{*}\}_{P.B.}=-{\cal H}.                 \eqno(3.19)
$$
This is the so called Hamiltonian constraint.
Poisson bracket between $\,C\,$ and $\,{\cal H}^{*}\,$ or $\,{\cal H}\,$
vanishes, so that we have no additional constraint.

By the canonical transformation which interchanges the coordinate ${\bf Q}$ 
and the momentum ${\bf P}$, the final form of $\,{\cal H}\,$ is obtained
$$
     {\cal H}={1\over2}{\bf Q}(\psi-{4\over3}{\bf P}^2-6k/a^2)
              -{1\over3}ap{\bf P}-\sqrt{h}f(\psi)                   \eqno(3.20)
$$
where $\,\psi=\psi({\bf Q}/2\sqrt{h})\,$.
Eq.(3.20) is obtained from (2.34) by using the relations ${\bf R}=6k/a^2$, 
$p'^{ij}=aph^{ij}/2d$ and putting $d=3$\\

\noindent
{\bf \S4. Summary and discussion}\\

Following the method of BL, we presented the canonical formalism of HCG
in which the Lagrangian density is given by $\,{\cal L}=\sqrt{-g}f(R)\,$.
Using the second class constraints, we reduced the phase space which is 
introduced naturally in the BL formalism.
At the last step of the procedure, we made a canonical transformation which 
interchanges the coordinate corresponding to the higher derivatives of 
the original coordinate and its conjugate momentum.
This transformation is necessary to avoid the non-locality when we proceed to 
the quantum theory.
In addition, comparison of the expressions for $\,\bar{\cal H}\,$, (2.32) with
(2.34), shows that the dependence of the Hamiltonian on the coordinates and 
the momenta becomes more natural by this transformation.


The dynamical degrees of freedom of the HCG considered here is just that of 
the Einstein gravity coupled to one scalar field.
The expression of $\bar{\cal H}$ in (2.34), however, does not appear to 
describe such a system.
We introduce a scalar field $\Phi$ and make a conformal transformation
$$
     \Phi\equiv \kappa^{-1}\ln{(\kappa^2{\bf Q}/2\sqrt{h})},\ \ \ \ \ 
     \bar{h}_{ij}\equiv \bigl(\kappa^2{\bf Q}/2\sqrt{h}\bigr)^{2/d}h_{ij}.
$$
If this is a part of a canonical trasnformation, corresponding canonical 
momenta $\Pi$ and $\bar{p}^{ij}$ are related to $\hat{\bf P}$ and $p'^{ij}$
as
$$
\left\{\begin{array}{l}
     \hat{\bf P}\dis ={\kappa^2\over \sqrt{\bar{h}}}\Bigl[\,
                       {1\over d}\bar{\bf p}+{1\over 2\kappa}\Pi\Bigr]
\\[5mm]
     p'^{ij}\dis=\exp{(2\kappa\Phi/d)}\Bigl[\bar{p}^{ij}-\Bigl({1\over d}
                 \bar{\bf p}+{1\over 2\kappa}\Pi\Bigr)\bar{h}^{ij}\,\Bigr].
\end{array}\right.
$$
Then $\bar{\cal H}$ takes the following form
$$
\begin{array}{ll}
     \bar{\cal H}&\dis ={\kappa^2\over \sqrt{\bar{h}}}\Bigl[\,
                           \bar{p}^{ij}\bar{p}_{ij}
                        -{1\over d-1}\bar{{\bf p}}^2\,\Bigr]
                        -\kappa^{-2}\bar{\bf R}
                      +{d-1\over 4d}{1\over \sqrt{\bar{h}}}
                            \Bigl[\Pi-{2\kappa\over d(d-1)}\bar{\bf p}\Bigr]^2
\\[5mm]
                &\dis +\kappa^{-2}\sqrt{\bar{h}}\Bigl[\psi
                       -{1\over2}\exp{(-\kappa\Phi)}F(\psi)
                       -\{\exp{(2\kappa\Phi/d)}-1\}\bar{\bf R}\Bigr]
\\[5mm]
                &\dis +{d+3\over d}\sqrt{\bar{h}}\exp{(2\kappa\Phi/d)}
                          \bar{h}^{ij}\del_{i}\Phi\del_{j}\Phi
\end{array}
$$
where $F\equiv f/2\kappa^2$.
This expression is very similar to the one for the Einstein gravity coupled 
with a scalar field.
The first two terms are just the Hamiltonian of the Einstein gravity.
However, the remaining terms are not the one of the scalar field used in 
the classical equivalence argument.
So, at this stage, the equivalence is not clear in the quantum theory, 
although there might exist a choice of variables which would show 
the equivalence explicitly.

For the Robertson-Walker spacetime, the second class constraint does not appear
since the phase space is reduced from the beginning.
This will make the situation simple when we consider the stability of 
the internal spaces.
In this case the second class constraints are restricted to those with respect 
to the internal variables.
Consequently the Dirac brackets are required only for these variables if 
necessary.
Investigation of this problem would require quantum mechanical treatment and
will be given in a separate paper.
\\
\vspace{8mm}

\noindent
{\bf Appendix. Dirac brackets}\\

In discussing the symmerty properties of constrained systems, Dirac bracket 
formalism is sometimes more convenient than the method using only 
the independent variables.
So we here list the Dirac brackets among the canonical variables.
$$
\begin{array}{l}
     \dis \{h_{ij}({\bf x}),p^{kl}({\bf x}')\}_{DB}
     =(3\delta_{i}^k\delta_{j}^l-{2\over d}h_{ij}h^{kl})\delta({\bf x-x'})
     \hspace{6.5cm}
\\[3mm]
     \{h_{ij}({\bf x}),h_{kl}({\bf x}')\}_{DB}=0
\\[3mm]
     \dis\{p^{ij}({\bf x}),p^{kl}({\bf x}')\}_{DB}
      =0
\\[3mm]
     \dis\{Q^{ij}({\bf x}),P_{kl}({\bf x}')\}_{DB}
     ={1\over d}h^{ij}h_{kl}\delta({\bf x}-{\bf x}')
\\[3mm]
     \{Q^{ij}({\bf x}),Q^{kl}({\bf x}')\}_{DB}=0
\\[3mm]
     \{P_{ij}({\bf x}),P_{kl}({\bf x}')\}_{DB}
      =0
\end{array}
$$
$$
\begin{array}{l}
     \{h_{ij}({\bf x}),Q^{kl}({\bf x}')\}_{DB}=0
\\[3mm]
     \dis\{h_{ij}({\bf x}),P_{kl}({\bf x}')\}_{DB}
     =2d{\bf Q}^{-1}(h_{ik}h_{jl}-{1\over d}h_{ij}h_{kl})\delta({\bf x-x'})
\\[5mm]
     \dis\{p^{ij}({\bf x}),Q^{kl}({\bf x}')\}_{DB}
     ={1\over d}{\bf Q}(h^{ik}h^{jl}-{1\over d}h^{ij}h^{kl})\delta({\bf x-x'})
\\[5mm]
     \dis\{p^{ij}({\bf x}),P_{kl}({\bf x}')\}_{DB}
     =\Bigl[{1\over d}(h^{ij}\tilde{P}_{kl}-\tilde{P}^{ij}h_{kl})
     -{\bf Q}^{-1}(2{\bf p}-{\bf PQ})(\delta_{k}^i\delta_{l}^j
        -{1\over d}h^{ij}h_{kl})\Bigr]\delta({\bf x-x'}).
\end{array}
$$
The Dirac brackets of $\,h_{ij},p^{kl}\,$ and $\,{\bf Q,P}\,$ are given as
$$
\begin{array}{l}
     \{{\bf Q},{\bf P}\}_{DB}=d\delta({\bf x}-{\bf x}')\hspace{11cm}
\\[3mm]
     \{h_{ij},{\bf Q}\}_{DB}=0
\\[3mm]
     \{h_{ij},{\bf P}\}_{DB}=0
\\[3mm]
     \dis\{p^{ij},{\bf Q}\}_{DB}=
                                  -{1\over d}h^{ij}{\bf Q}
                                   \delta({\bf x}-{\bf x}')
\\[5mm]
     \dis\{p^{ij},{\bf P}\}_{DB}=(2\tilde{P}^{ij}-{1\over d}h^{ij}{\bf P})
                                  \delta({\bf x}-{\bf x}')
\end{array}
$$

\end{document}